\input harvmac

\line{\hfil HUTP-98/A041}

\centerline{\bf Cosmological Searches for Photon Velocity Oscillations }
\smallskip

\centerline{\sl Sheldon L. Glashow }
\centerline{\sl Lyman Laboratory}
\centerline{\sl Harvard University}
\centerline{\sl Cambridge, MA 02138}\medskip

\centerline{\bf ABSTRACT}

{\smallskip\narrower\noindent We posit a second massless photon, uncoupled
to known forms of matter but undergoing Lorentz non-invariant velocity
mixing with ordinary photons. Our speculation within a speculation suffers
from the sin of implausibility but enjoys the virtue of verifiability. To
avoid unacceptable distortion of the well-measured microwave background
spectrum, the velocity difference  of the photons cannot exceed
$\sim\!10^{-32}\,c$.  Stronger constraints (or observable effects!) can
arise from  optical measurements of distant sources.\medskip}
\centerline{---}\medskip

Years ago we proposed the existence of a second photon to explain an
alleged spectral anomaly of the cosmic background radiation, which we 
attributed to mass mixing of  the two photons~\ref\ggg{H. Georgi,  S. L.
Glashow \&\ P. Ginsparg, Nature 306 (1983) 765.}.  However, the thermal
nature of the CBR has been established by COBE-FIRAS and the anomaly has
vanished~\ref\rcobe{D.J. Fixen,  E.S. Cheng, J.M. Gales, J.C. Mather, R.A.
Shafer \&\ E.L. Wright, {\tt http://xxx.lanl.gov/astro-ph/9605054.}}. Here
we revive the notion of a second photon $B$ with no couplings to ordinary
matter, but assume both $B$ and the usual photon $A$ to be massless.
Because $B$  decouples  earlier than $A$, its contribution to the entropy
of the universe is not significant. At this point, $B$ would have no
detectable effect. We proceed by conflating the notion of a second photon
with another speculation: the possible existence of tiny departures from
Lorentz invariance such as we have discussed elsewhere~\ref\rcg{ S. Coleman
\&\ S. L. Glashow, Phys. Lett. B405 (1997) 249.}.

We imagine our two photons to
experience Lorentz non-invariant velocity mixing.  
In a preferred
frame (presumably, the rest frame of the CBR), the velocity eigenstates
$a_1$ and $a_2$ 
propagate in any direction  at slightly different  velocities:
$v_1= c$ and $v_2=(1+\delta)\,c$. Should the interaction eigenstates 
of the photons differ from their velocity eigenstates:
$$\eqalign{
A\,{\rm (standard\ photon)}&= a_1\cos{\phi} +a_2\sin{\phi}\,,\cr
B\,{\rm (uncoupled\ photon)}&= a_2\cos{\phi} -a_1\sin{\phi}\,,\cr
}$$
ordinary photons would  oscillate in a fashion akin to the velocity
oscillations of  neutrinos~\rcg\ref\rmore{S.L. Glashow {\it et al.,} \
Phys. Rev. D56 (1997) 2433.}. Alternatively,  velocity oscillations of
photons could result from tiny violations of the equivalence principle in a
Lorentz invariant theory~\ref\rvep{M. Gasperini, Phys. Rev. D38 (1988)
2635;  A. Halprin and C.N. Leung, Phys. Rev. Lett. 67 (1991) 1833.}.

Photons emitted by ordinary matter start as $A$ but their
amplitude evolves as:
$$ a_1\,\cos{\phi} + a_2\,\sin{\phi} \,e^{-i\delta \,Et/\hbar}\,,$$
so that a non-interacting $B$ component develops with time.
The probability for an ordinary photon to remain an ordinary photon 
oscillates with time according to:
\eqn\eosc{P(t)= 1-b^2\,\sin^2{(\delta \,\omega t/2)}\,}
with $\omega$ the frequency of the detected photon and 
$b^2\equiv \sin^2{(2\phi)}$. 
We are concerned with the large mixing ($b\sim 1$) and small $\delta$
domain. Lorentz-violating differences between the velocity of light and the
maximum attainable velocities of massive bodies or particles have been
searched for in several contexts. The best current limits are of order
$10^{-22}c$~\ref\rlam{S. K. Lamoreaux, J. P. Jacobs,  B. R. Heckel, F. J.
Raab, and E. N. Fortson, Phys. Rev. Lett. 57 (1986) 3125.}\rcg. By
studying how photon
velocity oscillations affect radiation from sources at
cosmological distance, we constrain the Lorentz-violating parameter $\delta$ 
to be ten orders of magnitude smaller than the above limit. 

The effect of velocity oscillations on light from a source at redshift $z$
involves an integral over travel time of a redshifted signal.
Thus we must replace $\omega t$ in Eq.\eosc\ by:
\eqn\erep{\omega\int_1^{1+z} y\,dy\,(dt/dy)\equiv
\omega\,\hat{z}/H_0\;,}
where $H_0=h \times 100\ {\rm  km/s}\cdot$Mpc. The monotone function
$\hat{z}(z)$ is determined by
Eq.\erep\ and the redshift-time relation:
$${d\,y\over y}=-H(y)\,dt=-H_0\,\left[(1-\Omega)y^2 +(\Omega-\Lambda)y^3
+\Lambda\right]^{1/2 }\,dt\;,$$
with $\Lambda$ the cosmological constant and $\Omega_m\simeq
\Omega-\Lambda$.
For the special case
$\Omega=1$ and  $\Lambda=0$ 
we find: $\hat{z}(z)= 2\,(1-1/\sqrt{1+z})$, but in any case 
$\hat{z}$ approaches $z$ when it is
small.
The extinction coefficient, expressed as a function of redshift, becomes:
\eqn\eoscc{P(z) =1-b^2\sin^2{  \left\{ \delta\,\omega
\,\hat{z}/2H_0\right\}}\;.}

We first discuss the effect of these oscillations
on the cosmic background
radiation (CBR) spectrum.
Let $u(\omega)$ denote
the unperturbed thermal spectrum. Unless $b$ is tiny, the
oscillation length of CBR photons must exceed the horizon size
and the sine in Eq.\eoscc\ may be approximated by its argument.
The resulting spectral distortion is:\footnote{$^*$}{For the CBR in
a flat universe with $\Lambda=0$, it is appropriate to use
$\hat{z}(1100)\simeq 2$. 
For a flat universe with a positive cosmological 
constant, $\hat z$ (and the distortion) is
larger.}
$$ \delta u(\omega)\bigg\vert_{\rm osc} = -\left({b\,\delta \omega\over H_0}
\right)^2\,u(\omega)=
-{b^2\,\delta^2\, \omega^5\over \pi^2\, H_0^2 
\,(e^{\omega/\omega_0}-1)}\;,$$
where $\omega_0=3.57\times10^{11}$~s$^{-1}$ corresponds to the observed CBR
temperature.
 The oscillation-induced  distortion (which is largest at 
$\omega\simeq 5\,\omega_0$)
cannot exceed the observational 
limit on departures from a thermal spectrum~\rcobe:
$$\vert \delta u(\omega)\vert < 7.2\times10^{-6}\,\omega_0^3\;,
\quad\ {\rm for}\quad\ \omega_0<\omega<10\,\omega_0\;. $$
It follows that the oscillation parameters must satisfy:
\eqn\econst{b\,\delta/h< 1.6\times 10^{-32}\;.}
Departures from special relativity (and {\it a fortiori,} from general
relativity) would seem incompatible with our use of conventional cosmology.
Our  extraordinarily  strong constraint on $\delta$ ameliorates
but does not remove this
logical flaw.

Velocity oscillations can satisfy the above CBR constraint
and yet produce observable effects elsewhere because
visible light, with much greater frequency than the CBR, has a
correspondingly shorter  oscillation length. 
While Eq.\econst\ ensures that
there can be no detectable effect on nearby optical sources with $z<0.01$,
velocity oscillations could
systematically distort the spectra of more distant sources 
and change their apparent magnitudes.
The modulating function is $P(\lambda,\, z)= 1-b^2
\sin^2{\Phi}$,  where:
$$\Phi\simeq \hat{z}\times
11.6\,\left[{400\ {\rm nm}\over \lambda}\right]\,\left[
{h^{-1}\,\delta\over 1.6\times10^{-32}}\right]\;.$$
This modulation, if present, would complicate the determination of
cosmological parameters via analyses~\ref\rgold{The Supernova Cosmology
Collaboration:  S. Perlmutter {\it et al.,} ApJ 483 (1997) 565 and Nature
391 (1998) 51; G. Goldhaber and S. Perlmutter, Talks at Dark Matter 98,
Marina del Rey CA, Feb. 18, 1998.}\ref\rkir{The High-z Supernova Team:
P.M. Garnavich {\it et al.,} ApJ 493 (1998) L53; A.V. Filippenko, Talk at
Dark Matter 98, Marina del Rey CA, Feb. 18, 1998.} of redshifts and
apparent magnitudes of distant supernov\ae, conceivably reconciling these
data with the ``standard'' $\Omega=1$ and $\Lambda =0$ universe (at the
cost of adopting two rather implausible hypotheses). More likely, these
analyses will provide more stringent constraints on the oscillation
parameters. 

\medskip
Conversations with S. Coleman, H. Georgi, G. Goldhaber, S. Perlmutter, and
G. Smoot are gratefully acknowledged, as is the hospitality extended to me
by the Lawrence Berkeley National Laboratory and by S. Barish. This work
was partially supported by the National Science Foundation under 
grant NSF-PHYS
92-18167.
\listrefs
\bye